\documentclass[prl,aps,twocolumn,showpacs]{revtex4}

\usepackage{epsfig}

\begin{document}

\title{Simultaneous optimization of spin fluctuations and superconductivity 
under pressure in an iron-based superconductor}

\author{G. F. Ji$^{1}$}
\author{J. S. Zhang$^{2}$}
\author{Long Ma$^{1}$}
\author{P. Fan$^{1}$}
\author{P. S. Wang$^{1}$}
\author{J. Dai$^{1}$}
\author{G. T. Tan$^{3}$}
\author{Y. Song$^{3}$}
\author{C. L. Zhang$^{3}$}
\author{Pengcheng Dai$^{3,4}$}
\author{B. Normand$^{1}$}
\author{Weiqiang Yu$^{1}$}
\email{wqyu_phy@ruc.edu.cn}
\affiliation{
$^{1}$Department of Physics, Renmin University of China, Beijing 100872, China\\
$^2$School of Energy, Power and Mechanical Engineering, North China Electric 
Power University, Beijing 102206, China\\
$^{3}$Department of Physics and Astronomy, The University of Tennessee, 
Knoxville, Tennessee 37996-1200, USA\\
$^{4}$Beijing National Laboratory for Condensed Matter Physics, Institute 
of Physics, Chinese Academy of Sciences, Beijing 100190, China}

\date{\today}


\begin{abstract}

We present a high-pressure NMR study of the overdoped iron pnictide 
superconductor NaFe$_{0.94}$Co$_{0.06}$As. The low-energy antiferromagnetic 
spin fluctuations in the normal state, manifest as the Curie-Weiss upturn in 
the spin-lattice relaxation rate $1/^{75}T_1T$, first increase strongly with 
pressure but fall again at $P > P_{\rm opt} =$ 2.2 GPa. Neither long-ranged 
magnetic order nor a structural phase transition is encountered up to 2.5 GPa. 
The superconducting transition temperature $T_c$ shows a pressure-dependence 
identical to the spin fluctuations. Our observations demonstrate that magnetic 
correlations and superconductivity are optimized simultaneously as a function 
of the electronic structure, thereby supporting very strongly a magnetic origin 
of superconductivity.   

\end{abstract}

\maketitle

In the iron-based superconductors \cite{Hosono_Jacs_130_3296, 
Chen_PRL_100_247002, Chen_Nature_453_761, Ren_CPL_12_105}, charged dopants 
usually act to suppress an orthorhombic ground state with antiferromagnetic 
long-range order (AFLRO) in favor of a tetragonal, paramagnetic, and 
superconducting phase. Multiple electron bands are observed \cite{Lu_NaFeAs, 
Ye_PRX_2013}, which may include all five Fe $d$-orbitals. These results 
indicate that different electronic degrees of freedom, both orbital and 
magnetic, are involved in the fluctuations and possible broken-symmetry 
phases, and to date these complex correlation effects have obscured the 
pairing mechanism \cite{Paglione_Nature}. While spin fluctuations are a 
leading candidate for mediating superconductivity, orbital fluctuations 
have also been proposed for this role \cite{kontani_PRL_104_157001}. Direct 
evidence for the pairing mechanism continues to be the primary goal of the 
many studies investigating how the lattice structure, band structure, and 
magnetism determine the superconducting properties.

An applied pressure is a particularly clean method for controlling the 
physical properties of iron-based superconductors. The superconducting
transition temperature $T_c$ has been found to change strongly with 
pressure in LaFeAsO$_{1-x}$F$_x$ (1111 structure) \cite{Nakano_1111_P}, 
BaFe$_2$(As$_{1-x}$P$_x$)$_2$ (122) \cite{Ishida_pressure}, 
NaFe$_{1-x}$Co$_{x}$As (111) \cite{Wang_pressure}, Fe$_{1+x}$Se (11) 
\cite{Imai_prl_102_177005}, and many other systems \cite{Chu_review}. 
To date, NaFe$_{1-x}$Co$_x$As has shown the most marked effects, even of 
rather moderate pressures, in its structural, magnetic, and superconducting 
properties. NMR studies of the parent compound NaFeAs show that the N\'eel 
temperature $T_{\rm N}$ increases with pressure up to 2.4 GPa \cite{MaNaFeAs}, 
and x-ray measurements find a collapsed tetragonal phase above 3 GPa 
\cite{Liu_pressure}. These observations leave open the question of how 
changes in $T_c$ may be associated with competing spin fluctuations, 
AFLRO, and/or changes in crystal structure, and suggest a systematic 
study of correlation and pairing effects by changing the lattice 
parameters under pressure.

In this letter, we present a high-pressure $^{75}$As NMR study on the 
overdoped iron-based superconductor NaFe$_{1-x}$Co$_{x}$As with $x = 0.06$.
In the normal state, the spin-lattice relaxation rate divided by 
temperature, $1/^{75}T_1T$, first grows significantly with pressure, 
showing a low-temperature Curie-Weiss upturn indicative of strongly 
enhanced low-energy spin fluctuations. However, $1/^{75}T_1 T$ reaches 
a maximum at $P_{\rm opt} \simeq$ 2.17 GPa before decreasing again, a 
non-monotonic pressure-dependence not previously observed in iron-based 
superconductors. The superconducting transition temperature has an 
identical ``dome'' feature under pressure, with a maximal $T_c$ at the 
same $P_{\rm opt}$. These observations indicate clearly the strong correlations 
between magnetism, superconductivity, and the details of the underlying 
lattice, are quite different from the effects of doping, and give strong 
support for a magnetic origin of superconductivity. 

NaFe$_{1-x}$Co$_{x}$As is optimally doped at $x = 0.03$, where the 
maximal $T_c$ is approximately 20 K \cite{Chen_NaFeCoAs}. We perform a 
systematic study of pressure effects on the structure and the magnetic 
fluctuations, and of their correlation with superconductivity, by avoiding 
both the structural and magnetic phase transitions; for this we focus on a 
sample with significant overdoping, $x = 0.06$, where $T_c \approx$ 18 K. 
NaFe$_{0.94}$Co$_{0.06}$As single crystals were synthesized by the flux-grown 
method with NaAs as the flux. The doping was determined accurately from 
inductively coupled plasma atomic emission spectroscopy measurements. 

For our high-pressure NMR measurements we used a clamp-type pressure cell 
with Daphne oil as a pressure medium achieving high homogeneity. The clamp 
cell is limited to 2.5 GPa at low temperatures, and while $P$ cannot be changed 
externally below room temperature, it does change with $T$. For a complete 
calibration of the pressure at different temperatures we used a manometer 
of Cu$_2$O, whose nuclear quadrupole resonance (NQR) frequency is known 
very accurately \cite{Reyes_Cu2O}. We deduced $P(T)$ from $^{63} \nu_q (P,T)$, 
finding a pressure drop $\Delta P(T) \le 0.15$ GPa from 300 K to 150 K and 
negligible changes below 150 K. The pressures reported here are those we 
measured at 2 K. We stress that all our measurements under pressure were 
fully and reproducibly reversible. The superconducting transition under 
pressure was determined consistently by NMR and from the a.c.~susceptibility. 
The $^{75}$As NMR spectra were obtained by the spin-echo technique under a 
field of 7.63 T applied in the $ab$-plane. The spin-lattice relaxation rate 
$1/^{75}T_1$ was measured by the spin-inversion method. 

\begin{figure}
\includegraphics[width=8.4cm, height=6.2cm]{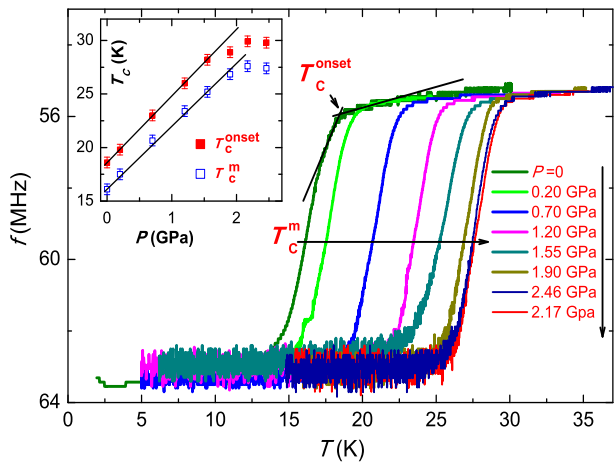}
\caption{\label{acsus}(color online) Main panel: RF resonance frequency 
of the detuned NMR circuit measured as a function of temperature and 
pressure at zero field. The onset and mid-point superconducting transition 
temperatures, respectively $T^{\rm onset}_c$ and $T^{\rm m}_c$, are indicated 
by the arrows. Inset: values of $T^{\rm onset}_c$ and $T^{\rm m}_c$ as functions 
of pressure.}
\end{figure}

$T_c$ can be determined accurately {\it in situ} at all pressures by 
the a.c.~inductance change of the sample coil during cooling and warming 
at zero field. The superconducting transition is indicated (Fig.~\ref{acsus}) 
by an increase in the resonance frequency of the NMR circuit, which measures 
the a.c.~susceptibility, upon cooling. We define the onset ($T^{\rm onset}_c$) 
and mid-point ($T^{\rm m}_c$) temperatures from the frequency curve, as 
illustrated in Fig.~\ref{acsus}. Both $T^{\rm onset}_c$ and $T^{\rm m}_c$ have 
a strong initial increase (6 K/GPa) with pressure (inset, Fig.~\ref{acsus}). 
However, after reaching maximal values of 29.8 K ($T^{\rm onset}_c$) and 27.4 K 
($T^{\rm m}_c$) at a pressure $P_{\rm opt} \simeq$ 2.17 GPa, both quantities 
then decrease slowly ($-$ 0.6 K/GPa) at higher pressures. This dome-shaped 
superconducting transition is consistent with the results of high-pressure 
transport studies \cite{Wang_pressure}.    

\begin{figure}
\includegraphics[width=8.4cm, height=6.2cm]{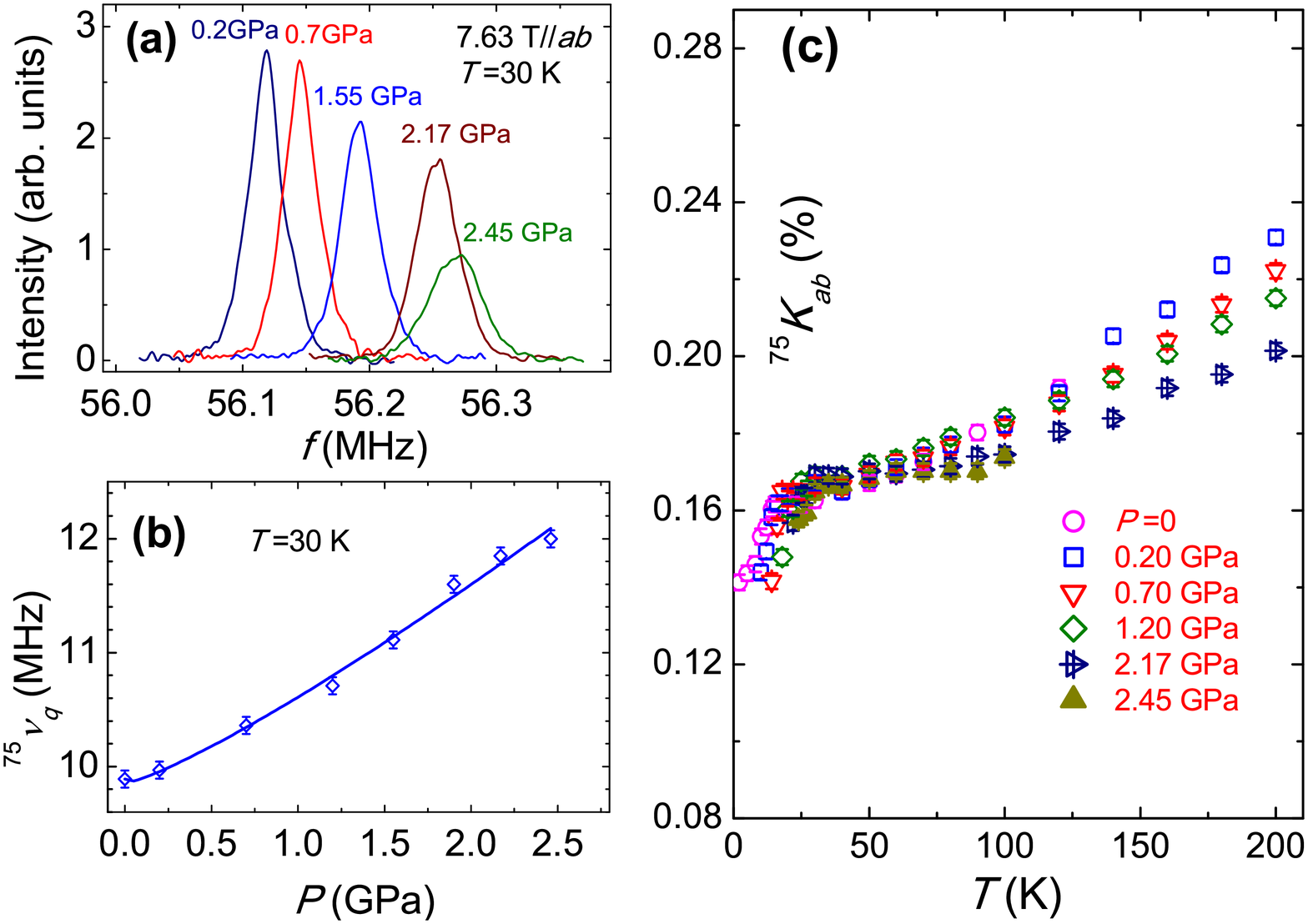}
\caption{\label{kvsp}(color online) (a) Center line of the $^{75}$As NMR 
spectra at different pressures, with field applied in the crystalline 
$ab$-plane. (b) Pressure-dependence of the $^{75}$As quadrupole frequency 
$^{75}\nu_q$ at $T =$ 30 K. (c) Temperature-dependence of the Knight shift 
$^{75}K_{ab}$ at different pressures.}
\end{figure}

We have measured $^{75}$As ($S = 3/2$) NMR spectra over the full temperature 
range to 200 K, at a number of different pressures and with the field applied 
in the $ab$-plane. Figure \ref{kvsp}(a) shows the temperature-corrected center 
line of the spectrum at $T = 30$ K for several pressure values. The spectra 
shift monotonically to higher frequencies, primarily as a result of 
second-order corrections from the $^{75}$As quadrupole frequency, $^{75}\nu_q$, 
which we discuss below. The NMR line width increases from 25 kHz at $P =$ 0 
to 50 kHz at $P =$ 2.46 GPa, showing a weak pressure inhomogeneity at higher 
pressures. 

The quadrupole frequency is measured from the $^{75}$As satellite spectra 
(data not shown). The low-temperature values of $^{75}{\nu}_q$ display an 
appreciable rise with pressure up to 2.46 GPa [Fig.~\ref{kvsp}(b)]. 
$\nu_q$ measures the local electric field gradient (EFG), which is very 
sensitive to the lattice parameters. This continuous increase of $^{75}\nu_q$ 
indicates a continuous lattice compression under pressure; neither the line 
shape nor the satellite frequency shows any abrupt changes with pressure or
temperature. Thus the structure remains tetragonal and a transition to 
orthorhombic or collapsed-tetragonal symmetry is excluded up to 2.46 GPa, 
in contrast to the behavior observed in NaFeAs \cite{MaNaFeAs, Liu_pressure}.  

The in-plane Knight shift $^{75}K_{ab}$ deduced from the center line of the 
NMR spectrum is shown in Fig.~\ref{kvsp}(c). At a fixed pressure, $^{75}K_{ab}$ 
increases monotonically with temperature; the functional form $^{75}K_{ab} = 
A_0 + B_0 T + C_0 T^2$ is characteristic of additive contributions from 
itinerant electrons ($A_0$) and from predominantly two-dimensional (2D) local 
spin fluctuations ($B_0$) \cite{Ma_prb_84}, with only weak 3D contributions 
from inter-plane coupling ($C_0$). There are no abrupt changes in $^{75}K_{ab}$;
taken together with constant Boltzmann-corrected spectral intensities down to 
1.5 K at each pressure and the absence of diverging behavior in $1/^{75}T_1$ 
above $T_c$ (shown below), this excludes a magnetic ordering transition below 
2.46 GPa. At a fixed temperature $T > T_c$, $^{75}K_{ab}$ decreases with pressure.
At $T < T_c$, $^{75}K_{ab}$ drops sharply, indicating a singlet superconducting 
order parameter. The values of $T_c$ determined from the Knight shift are fully 
consistent with those from the a.c.~susceptibility data (Fig.~\ref{acsus}). 

\begin{figure}
\includegraphics[width=8.4cm, height=6.2cm]{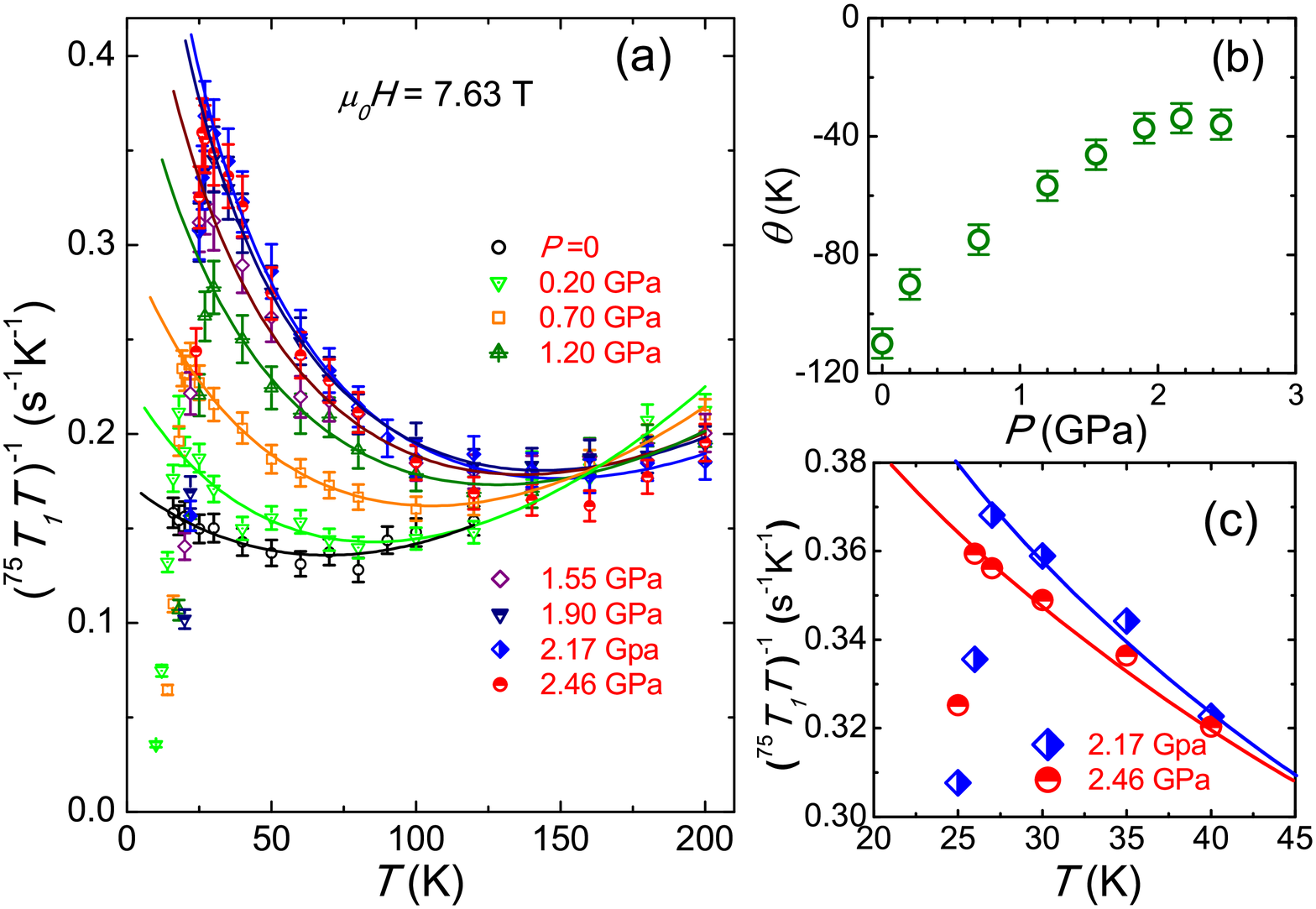}
\caption{\label{t1vstp}(color online) (a) Temperature-dependence of 
$1/^{75}T_1 T$ at different pressures. The solid lines are fits to the form
$1/^{75}T_1 T = A_1/(T - \theta) + B_1 T + C_1 T^2$. (b) Pressure-dependence 
of the Curie-Weiss temperature $\theta$ extracted from panel (a). 
(c) Comparison of $1/^{75}T_1 T$ data near $T_c$ at the two highest 
pressures [data and fitting lines as in panel (a)].}
\end{figure}

The $^{75}$As spin-lattice relaxation rates ($1/^{75}T_1$) measured at 
each pressure are shown in Fig.~\ref{t1vstp} (a) for temperatures up to 
200 K. On cooling, $1/^{75}T_1 T$ first decreases but then shows a broad, 
low-temperature upturn before falling abruptly below $T_c$. The upturn, 
which becomes increasingly prominent at high pressures, can be fitted 
rather well by the expression $1/^{75}T_1 T = A_1/(T - \theta) + B_1 T + C_1 
T^2$. The Curie-Weiss contribution ($A_1$) is consistent with 2D low-energy 
spin fluctuations \cite{Moriya}, and demonstrates their increasing importance 
as pressure drives the system closer to a magnetic ordering transition.
However, unlike underdoped NaFe$_{1-x}$Co$_x$As, where $1/T_1 T$ diverges 
at the onset of AFLRO \cite{MaNaFeAs}, our overdoped sample shows no 
divergence. Instead, the values of $|\theta|$ extracted from the fit 
at each pressure, shown in Fig.~\ref{t1vstp}(b), approach the divergent 
regime but then increase again. We stress that $1/^{75}T_1 T$ at low 
temperatures shows the same non-monotonic pressure-dependence as $T_c$ 
[Fig.~\ref{t1vstp}(c)], {\it i.e.}~the low-energy spin fluctuations are 
optimized at the same pressure $P_{\rm opt}$. This behavior is also reflected 
in the maximum of $\theta$ [Fig.~\ref{t1vstp}(b)], which maximizes the 
Curie-Weiss term.

\begin{figure}
\includegraphics[width=8.4cm, height=6.2cm]{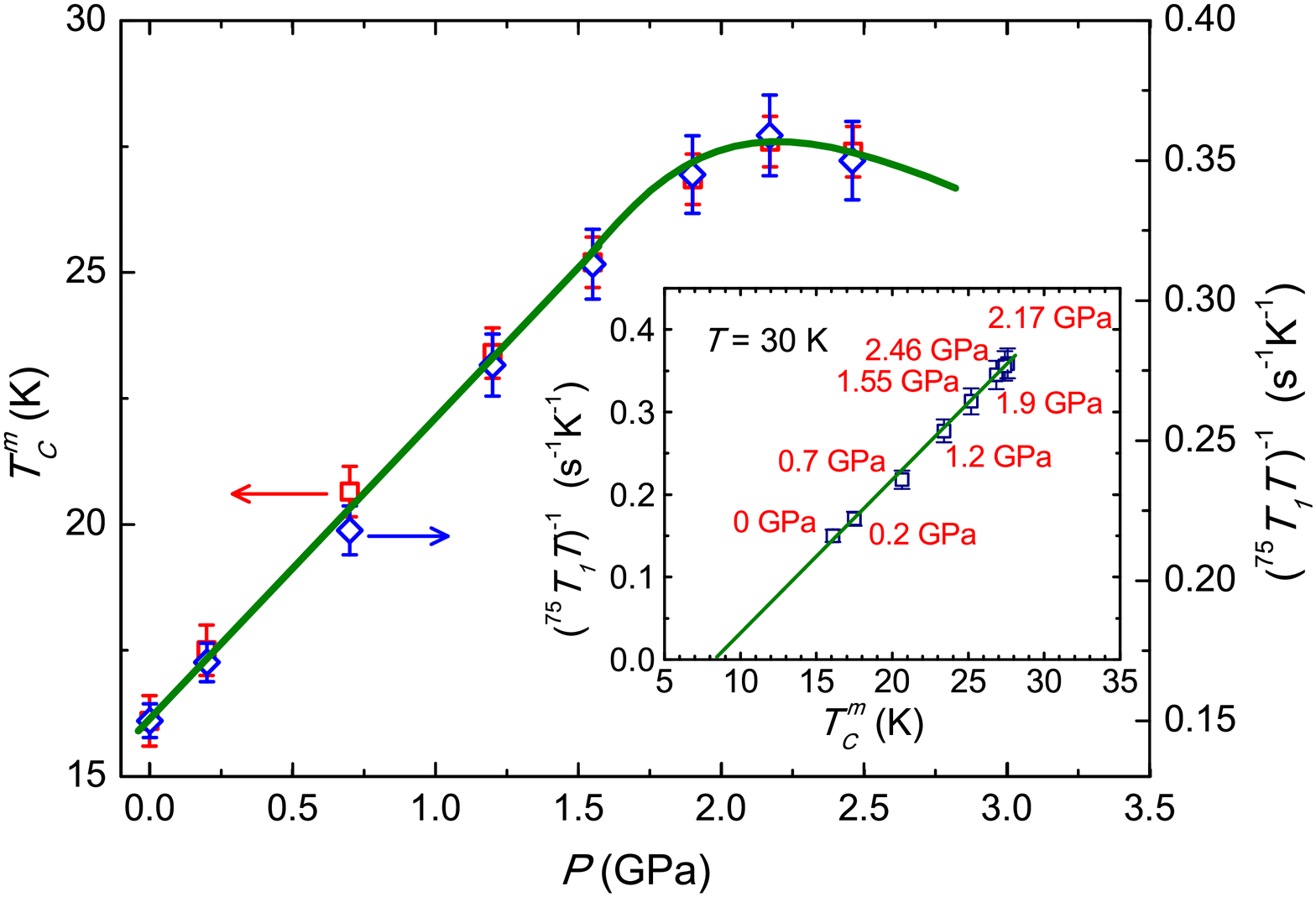}
\caption{\label{tcvst1}(color online) Main panel: mid-point superconducting 
transition temperature $T^{\rm m}_c$ (squares) and normal-state spin-lattice 
relaxation rate $1/^{75}T_1 T$ at $T =$ 30 K (diamonds) as a function of 
pressure. Inset: scaling between $T_c$ and normal-state $1/^{75}T_1 T$.}
\end{figure}

We conclude our data analysis by performing a detailed comparison between 
$T_c$ and the low-energy spin-fluctuation contribution to $1/^{75}T_1 T$. 
Figure \ref{tcvst1} shows $1/^{75}T_1 T$ at $T =$ 30 K, directly above $T_c$, 
and $T^{\rm m}_c$ taken from Fig.~\ref{acsus}, for all measured pressure 
values. The two quantities have an initial linear increase, begin to 
flatten above 1.7 GPa, are maximal at 2.17 GPa, and fall beyond this. 
To our knowledge, such a simultaneous optimization of $T_c$ and the 
low-energy spin fluctuations in an unconventional superconductor has not 
been demonstrated before. We have achieved this optimization through the 
pressure-dependence of both quantities while avoiding the structural and 
magnetic phase transitions. To make the relationship between magnetic 
fluctuations and superconductivity yet more explicit, in the inset of 
Fig.~\ref{tcvst1} we plot $1/^{75}T_1 T|_{T = 30 \; {\rm K}}$ against $T_c$ 
with pressure as the implicit parameter. The pressure-induced changes 
$\Delta (T_c)$ and $\Delta (1/^{75}T_1 T|_{T = 30 \; {\rm K}})$ show a simple 
linear scaling behavior, valid both below and above the optimal pressure.

We begin our discussion by considering the low-energy spin fluctuations. 
Irrespective of the connection to superconductivity, such an optimization
of spin fluctuations by changing the lattice parameters has also not been 
observed previously. This non-monotonic change clearly cannot be described 
by any sort of effective (negative) doping, because doping always leads to 
AFLRO in Fe-based superconductors, with the dome of optimal doping arising 
due to the competition between magnetic order and superconductivity. The 
behavior we observe also contrasts strongly with the effects of pressure 
in FeSe, where spin fluctuations increase monotonically until AFLRO sets in 
\cite{Imai_prl_102_177005}.

Because the spin fluctuations can be optimized by pressure without a change 
of structural symmetry, our results demonstrate that the magnetic interactions 
are extremely sensitive to the exact lattice parameters, and therefore supply 
information important for a microscopic model. Although the iron-based 
superconductors have a complex, multi-orbital electronic structure, the 
Fermi surfaces of NaFe$_{1-x}$Co$_{x}$As and their orbital composition have 
been well characterized by Angle Resolved Photoemission Spectroscopy 
(ARPES) \cite{Lu_NaFeAs, Liu_NaFeCoAs,Ye_PRX_2013}. NaFe$_{1-x}$Co$_{x}$As is 
a quasi-2D system whose band structure is only weakly dispersive along the 
$c$-axis. Under these circumstances, one expects that the primary effect 
of an applied pressure will be to compress the $c$-axis lattice parameter; 
this interpretation is consistent with the large but continuous increase 
of $^{75}\nu_q$ [Fig.~\ref{kvsp}(b)], which is determined by $V_{zz}$, 
the principal EFG in the tetragonal phase. Because the As sites lie 
above and below the Fe layers, $c$-axis compression increases the overlap 
between the Fe $d_{xz}$- and $d_{yz}$-orbitals and the As $p$-orbitals. The 
pressure-enhanced low-energy spin fluctuations should thus be associated 
with improved Fermi-surface nesting of the $d_{xz}$ and $d_{yz}$ orbitals, 
a result confirmed by a recent study combining ARPES and NMR measurements 
on NaFe$_{1-x}$Co$_{x}$As \cite{Ye_PRX_2013}. 

However, the decrease in spin fluctuations beyond $P_{\rm opt}$ raises 
further questions. High-pressure synchrotron x-ray powder diffraction 
studies of NaFeAs found that the FeAs planes achieve a structure where 
the FeAs$_4$ tetrahedra are completely regular (all internal angles equal 
to 109.4$^{\rm o}$) at approximately 3 GPa \cite{Liu_pressure}. This regular 
structure appears to optimize the superconducting transition temperature in 
many iron pnictides \cite{optimalangle,Zhao_NM}. Although we cannot probe the 
lattice structure by NMR, our results for Co-doped NaFeAs certainly display 
a similar optimization as a function of lattice distortion, presumably as 
the ``horizontal'' and ``vertical'' As-Fe-As bond angles approach the regular 
value from opposite directions under pressure. Our data therefore imply that 
the empirical observation of a maximal $T_c$ and the achievement of completely 
regular FeAs$_4$ tetrahedra \cite{optimalangle,Zhao_NM} may be connected by 
the optimization of magnetic correlations. A possible origin for this effect 
could lie in the optimization of Fermi-surface nesting.

Considering the spin fluctuations in more detail, our data show that they 
have two different types in NaFe$_{1-x}$Co$_{x}$As. One is the low-energy spin 
fluctuations, responsible for the Curie-Weiss upturn at low temperatures in 
$1/^{75}T_1 T$. These usually arise due to itinerant electrons and are observed 
both by ARPES \cite{Ding_EPL_83_47001} and by INS \cite{Dai_NP_8_708} in 
compounds with good Fermi-surface nesting; they are peaked at the wave vector 
of the incipient AFLRO, and hence dominate $1/^{75}T_1 T$ \cite{Ning_PRL_104, 
Kita_JPSJ_77_114709} but are scarcely evident in the Knight shift. However, 
this upturn is weak in overdoped 1111 materials \cite{Nakano_1111_P}, 
completely absent in the intercalated iron selenide K$_y$Fe$_{2-x}$Se$_2$ 
\cite{YuW_11011017}, and weak in NaFe$_{0.94}$Co$_{0.06}$As at ambient pressure 
[Fig.~\ref{t1vstp}], and yet these systems all have a high $T_c$. To identify 
the origin of strong pairing interactions in these compounds, we note that 
their Knight shifts increase significantly with temperature, as observed 
respectively in Ref.~\cite{Nakano_1111_P}, Ref.~\cite{Ma_prb_84}, and 
Fig.~\ref{kvsp}(c). In fact this strong thermal enhancement appears in both 
$^{75}K_{ab}$ and $1/^{75}T_1 T$, meaning at all wave vectors, and its functional 
form [the relative linear ($B_0$, $B_1$) and quadratic ($C_0$, $C_1$) 
coefficients] is consistent with other indicators of predominantly 2D or 
3D nature. This behavior is characteristic of fluctuating local moments 
\cite{Ma_prb_84}, rather than itinerant electrons and a band-structure 
description \cite{Ikeda_JPSJ}. Our data show that the low-energy spin 
fluctuations are strongly enhanced by the pressure [Fig.~\ref{t1vstp}], 
while the local spin fluctuations are strongest at low pressures but weaken 
as $P$ increases [Fig.~\ref{kvsp}(c)].

Turning now to the connection with superconductivity, the paradigm of 
a spin-fluctuation-mediated pairing interaction whose strength diverges 
at the magnetic instability in the random phase approximation (RPA) was 
the foundation for several theories of high-temperature superconductors. 
However, in cuprates the separation in doping between the AFLRO phase 
and the dome-shaped maximum in $T_c$ is impossible to reproduce in this 
scenario. Here we obtain a direct proof for the correlation between 
low-energy spin fluctuations and superconductivity by their simultaneous 
optimization, using pressure as the control parameter. This is a very 
strong statement in favor of a magnetic origin for superconductivity. We 
reiterate that the pressure-enhanced $T_c$ we observe is correlated more 
directly with the low-energy spin fluctuations, caused by itinerant 
electrons, than with the local ones. This behavior is also manifest in 
the doping-dependence of the two spin-fluctuation types, where the 
high-energy ones were found \cite{Liu_NP} to change little with electron 
doping in BaFe$_2$As$_2$ while significant changes were found in the 
low-energy ones.

Our observations also shed light on the question of whether superconductivity 
in iron-based materials requires low-energy spin fluctuations at all, given 
that these seem to be weak or absent in some systems. By monitoring the 
evolution of NMR response with pressure, we have shown how superconductivity 
is correlated with two types of spin fluctuation. To distinguish between their 
contributions, we note in the perfectly linear relation between $1/^{75}T_1 T$ 
and $T_c$ (inset, Fig.~\ref{tcvst1}) that $T_c$ extrapolates to a finite value 
(around 8 K) as $1/^{75}T_1 T \rightarrow 0$. This indicates that low-energy 
spin fluctuations are not the only contribution to pairing, and that 
superconductivity may arise in their absence. Given the presence of local 
spin fluctuations, which are strong at low pressures [Fig.~\ref{kvsp}(c)], 
we suggest that these are the short-range magnetic correlation effects 
providing the additional pairing interaction, which is dominant in some 
materials. In NaFe$_{0.94}$Co$_{0.06}$As, our data show both local and low-energy 
spin fluctuations contributing to superconductivity at ambient pressure, 
while the latter dominate at high pressures; this balance of contributions 
will change with sample doping. 

Finally, spin fluctuations are not the only candidate pairing mechanism 
in Fe superconductors. Pairing mediated by orbital fluctuations has been 
proposed in a five-band model with electron-phonon coupling 
\cite{kontani_PRL_104_157001}. Our data resolve this question. The 
direct correlation of $T_c$ and $1/^{75}T_1 T$ favors unequivocally 
a magnetic origin. Further, phonon-mediated interactions are expected to 
increase monotonically with pressure, and so a non-monotonic change in 
$T_c$ does not appear to be consistent with the orbital-fluctuation 
scenario. A further consequence of this mechanism would be a conventional 
$s^{++}$ pairing symmetry, which should result in an NMR coherence peak 
robust against disorder. We are uniquely positioned to comment on the 
pairing symmetry, and we find that $1/^{75}T_1 T$ drops sharply below $T_c$, 
{\it i.e.}~the coherence peak is absent at all pressures. This result 
indicates an unconventional pairing symmetry such as $s^{+-}$, which is 
sensitive to impurity scattering \cite{Ishida_review}, again contradicting 
the orbital-fluctuation prediction. We found no evidence for a change of 
pairing symmetry under pressure. 

In summary, we have demonstrated a direct connection between 
superconductivity and low-energy spin fluctuations in a high-temperature 
superconductor. We chose to analyze NaFe$_{0.94}$Co$_{0.06}$As, an overdoped 
system where both the structural phase transition and antiferromagnetic 
long-range order are avoided. We performed NMR measurements under an 
applied pressure, which allows clean and detailed control of both the 
lattice and electronic structures. We show that the spin fluctuations and 
the superconducting transition temperature change in lockstep, and are 
optimized at exactly the same pressure. This result strongly supports a 
magnetic origin for superconductivity. Our measurements also demonstrate 
the presence of two types of spin fluctuation, namely low-energy ones 
arising from itinerant electrons and finite-energy ones with a local 
nature, and that both contribute to pairing in the superconducting state. 

The authors acknowledge helpful discussions with W. Bao, D. L. Feng, 
S. L. Li, and S. C. Wang. Work at Renmin University of China is supported 
by the National Science Foundation of China (Grant Nos.~11174365, 
11222433, and 11374364) and the National Basic Research Program of China 
(Nos.~2010CB923004, 2011CBA00112, and 2012CB921704). Work at North 
China Electric Power University is supported by the National Science 
Foundation of China (Grant No.~11104070). Work at the University of 
Tennessee is supported by the US DoE Office of Basic Energy Sciences 
under Grant No.~DE-FG02-05ER46202. 


\end{document}